\documentclass[conference]{IEEEtran}

\usepackage{graphicx}
\usepackage{epstopdf}
\usepackage{multirow}
\usepackage{bbm} 
\usepackage{amsfonts} 
\usepackage{amssymb}  
\usepackage{amsmath} 
\usepackage{url}
\usepackage{caption}
\usepackage{subcaption}
\usepackage{color}

\newtheorem{theorem}{Theorem}[section]
\newtheorem{lemma}[theorem]{Lemma}

\newtheorem{remark}[theorem]{Remark}

\newcommand*{\qed}{\hbox{}\hfill$\Box$}
\newcommand{\sst}{\scriptsize\mbox}

\newcommand{\rhou}{\rho_{\scriptsize\mbox{u}}}

\IEEEoverridecommandlockouts 

\begin{document}

\title{Can Massive MIMO Support\\Uplink Intensive Applications?}

\author{
\IEEEauthorblockN{Hong Yang}
\IEEEauthorblockA{\it Nokia Bell Labs, Murray Hill, NJ, USA\\
\tt{h.yang@nokia-bell-labs.com}}
\and
\IEEEauthorblockN{Erik G. Larsson*}
\IEEEauthorblockA{\it Link\"oping University, Sweden\\
\tt{erik.g.larsson@liu.se}}\thanks{*The work of E. G. Larsson was supported by the Swedish Research Council (VR) and ELLIIT.}
}

\maketitle

\begin{abstract}
Current outdoor mobile network infrastructure cannot support uplink
intensive mobile applications such as connected vehicles that collect
and upload large amount of real time data. Our investigation reveals
that with maximum-ratio (MR) decoding, it is theoretically impossible
to support such applications with cell-free Massive MIMO, and it
requires a very large number of service antennas in single cell
configuration, making it practically infeasible; but with zero-forcing
(ZF) decoding, such applications can be easily supported by cell-free
Massive MIMO with very moderate number of access points (AP's), and it
requires a lot more service antennas in single cell configuration. Via
the newly derived SINR expressions for cell-free Massive MIMO with ZF
decoding we show that uplink power control is unnecessary, and that
with 10 MHz effective bandwidth for uplink data transmission, in urban
and suburban morphologies, on the 2 GHz band, 90/$\mbox{km}^2$ and
32/$\mbox{km}^2$ single antenna AP's are enough to support 18
autonomous vehicles respectively. In rural morphology, using 450 MHz
band, only 2/$\mbox{km}^2$ single antenna AP's is enough.
\end{abstract}


\section{Introduction}
\IEEEPARstart{U}{nlike} traditional wireless applications which are
downlink intensive, many new IoT (Internet of Things) applications
require very high uplink data rates, ultra-low latency and high
reliability. While some applications, such as telesurgery, can be
supported by wireline transmission technologies such as optical fiber,
there are applications that must be supported wirelessly. For example,
connected vehicles, such as fully autonomous moving automobiles and
drones, may be required to continuously transmit large amounts of
real-time data collected by the many sensors about their surroundings
to the control center with high reliability and little delay. For
those applications, the required uplink data rate can be magnitudes
higher than what current 4G wireless technologies can support. For
example, each connected car can generate more than 25 gigabytes of
data every hour \cite{hitachi_2015}. To upload those data in real time
requires a sustained uplink data rate of about 56 Mbps per
vehicle. Supporting such high throughput in a highly mobile
environment requires very high spectral efficiency because the large
bandwidth in mmWave (millimeter wave) is not suitable for mobile
applications.

Can a Massive MIMO (multiple-input and multiple-output) wireless
network meet such a demand? We shall investigate the feasibility of
using MR (maximum-ratio) and ZF (zero-forcing) decoding to support
such high uplink throughput applications in both cellular and
cell-free configurations. Our investigation yields some surprises --
both pleasant and unpleasant ones.

\section{Massive MIMO System Model}\label{s:mMIMO}

Massive MIMO \cite{tlm_2010} utilizes a large number of service
antennas and recurrently updated CSI (channel state information) to
enable precise beamforming to a smaller number of user terminals. 
The service antennas can be co-located at base stations or distributed throughout the coverage area.

{\it Cellular Massive MIMO} divides the intended coverage area into
cells. Each cell is served by a base station with many co-located service
antennas. The base stations do not cooperate except for power control. In a multi-cell system, users in each cell is served by the cell's base
station antennas. Thus each user is served by all the service antennas
in the system only in single cell case.

{\it Cell-Free Massive MIMO} distributes many service antennas as access
points to provide data service throughout the entire intended coverage
area. Each user is served by all the service antennas in the system.

\subsection{Uplink Data Channel}
For a single cell or a cell-free Massive MIMO, the uplink data channel is modeled as 
$$
{\bf y}=\sqrt{\rhou}G{\bf s}+{\bf w}
$$
where
${\bf y}\in {\mathbb C}^M$ is the signal vector received by the $M$ service antennas; 
$\rhou >0$ is the normalized uplink SNR (signal-to-noise ratio);
$G\in {\mathbb C}^{M\times K}$ is the channel matrix between the $M$ service antennas and the $K$ mobile terminals, where $M >K$;
$
{\bf s}=\left[\sqrt{\eta_1}q_1, \cdots, \sqrt{\eta_K}q_K\right]^T \in {\mathbb C}^K
$ 
is the power controlled user message vector from the $K$ mobile terminals. Here 
\begin{equation}\label{eq:eta}
{\boldsymbol\eta}=\left[\eta_1, \cdots, \eta_K\right]^T \in [0,1]^K
\end{equation}
is the uplink power control vector. We assume that the user data symbols 
$
{\bf q}=\left[q_1, \cdots, q_K\right]^T \in {\mathbb C}^K
$
satisfies
$
{\text E}\left({\bf q}{\bf q}^* \right)= I_K,
$
where ${\text E}$ denotes expectation, and
${\bf w}\in {\mathbb C}^M$ is the noise vector, with each entry of the vector distributed as CN(0,1) and mutually independent. Here CN($\mu, \sigma^2$) denotes the circularly
symmetric Gaussian random variable with mean equals $\mu$ and variance
equals $\sigma^2$.

The channel between the $m$th service antenna and the $k$th user is modeled as 
$$
g_{m,k}=\sqrt{\beta_{m,k}}h_{m,k},
$$
where $\beta_{m,k}$ models the large-scale fading that accounts for geometric attenuation and shadow fading; $h_{m,k}$ models the small-scale fading that accounts for random scattering.
In a rich scattering propagation environment, the magnitude of the signal typically varies randomly according the Rayleigh distribution, thus $h_{m, k}, \forall m, k$ are modeled as independent and identically distributed CN(0,1) random variables.  

In single cell, the $M$ service antennas are co-located at a base station. The large-scale fading between the $k$th user and each of the $M$ service antennas is substantially the same, i.e.,
$$
\beta_{m,k} = \beta_k, \quad \forall m.
$$

\section{Uplink Performance}

In this section, we shall examine the possibilities of using MR and ZF
decoding to support uplink intensive wireless applications in both
single-cell and cell-free settings, where each user is served by all
the service antennas in the system.

Theoretical investigation accompanies a practical example to
demonstrate the performance differences among these four
configurations.

The example is to support 18 connected vehicles. Each vehicle must
upload 25 GB of data per hour \cite{hitachi_2015} continuously. To
upload this amount of data to the control center we need an uplink
throughput of about 56 Mbps for each vehicle. A Massive MIMO with TDD
(time division duplex) operation expend a portion of coherence
interval for uplink pilot, and the rest of coherence interval is
divided between downlink and uplink for data transmissions
\cite{mlyn_2016}. We assume a total spectral bandwidth of 20 MHz. If
we use half of the coherence interval for uplink data transmission,
then effectively 10 MHz is used for uplink data transmission, leaving
10 MHz for both uplink pilot and downlink data transmission. Thus 6
bps/Hz spectral efficiency will provide 60 Mbps uplink throughput. We
assume that each vehicle terminal has 2 Watts of available radiated
power. All the antenna gains are 0 dBi; the receiver noise figure is 9
dB. These parameters are used to calculate $\rhou$ as in Appendix F of
\cite{mlyn_2016}. Note that the total bandwidth of 20 MHz is used to
calculate the noise power. Simulations are carried out to determine
the required number of service antennas $M$ to achieve per vehicle
uplink spectral efficiency of 6 bps/Hz.

Using $\hat G$ and $\tilde G$ to denote the MMSE (minimum mean square
error) estimation and estimation error respectively, we have
\cite{mertins_1999}
\begin{equation}\label{eq:ghat}
[\hat G ̂]_{m,k}\sim \mbox{CN}\left(0,{\rhou\tau\beta_{m,k}^2\over 1+\rhou\tau\beta_{m,k}}\right)
\end{equation}
and the estimation error 
\begin{equation}\label{eq:gtilde}
[\tilde G ̂]_{m,k}\sim \mbox{CN}\left(0,{\beta_{m,k}\over 1+\rhou\tau\beta_{m,k}}\right)
\end{equation}
where $[A]_{m,k}$ denotes the entry at the $m$th row and $k$th column
of the matrix $A$, and $\tau$ denotes the length of mutually
orthogonal uplink pilot sequences that are used for channel
estimation.

\subsection{Simulation Parameters}
\subsubsection{Propagation Models}
Traditional Hata/COST231 propagation models are not suitable for
cell-free systems because the distances between transmitters and
receivers can be much shorter than the applicable range of these
models.  We use the ``NLoS'' propagation models specified in
\cite{ITU-R_2009}, for which the path loss in dB is given by
\begin{eqnarray*}
\mbox{PL}(d) =&&\hspace{-.25in} 161.04  - 7.1 \log_{10} (W) + 7.5 \log_{10} (h) \\
&&\hspace{-.25in}  -  [24.37 - 3.7(h/h_{\sst{AP}})^2]\log_{10} (h_{\sst{AP}}) \\
&&\hspace{-.25in} + [43.42 - 3.1 \log_{10} (h_{\sst{AP}})] [\log_{10} (d) - 3] \\
&&\hspace{-.25in} +20 \log_{10}(f_{\sst c}) - (3.2 [\log_{10} (11.75 h_{\sst{AT}})]^2 - 4.97)
\end{eqnarray*}
where $W$ is the street width (in meters); $h$ is the average building
height (in meters); $h_{\sst{AP}}$ is the access point antenna height
(in meters); $h_{\sst{AT}}$ is the user access terminal antenna height
(in meters); and $f_{\sst c}$ is the carrier frequency in GHz; $d$ is
the distance between transmitter antenna and receiver antenna, also in
meters.

The simulation parameters for the propagation models are compatible
with \cite{ITU-R_2009} and are summarized in Table \ref{t:spm}, where
$\sigma_{\sst{sf}}$ is the standard deviation of the lognormal shadow
fading.

\begin{table}[!t]
\renewcommand{\arraystretch}{1.3}  
\caption{Simulation Parameters}
\label{t:spm}
\centering{
\begin{tabular}{|c|c|c|c|}
\hline
& Urban & Suburban & Rural \\
\hline
$W$ (in meters) & 20  & 20 & 20 \\
\hline
$h$ (in meters) &  20 & 10 & 5 \\
\hline
$h_{\sst{AP}}$ (in meters) &  20 & 20 & 40 \\
\hline
$h_{\sst{BS}}$ (in meters) &  50 & 50 & 50 \\
\hline
$h_{\sst{AT}}$ (in meters) &  1.5 & 1.5 & 1.5 \\
\hline
$f_{\sst c}$ (in GHz)  &  2 & 2 &  0.45 \\
\hline
$\sigma_{\sst{sf}}$ (in dB)  &  6 & 8 &  8 \\
\hline
$R$ (in km) & 0.5  & 1 & 4 \\
\hline
$K$ & 18 & 18 & 18 \\
\hline
$M$ & ? & ? & ? \\
\hline
\end{tabular}
}
\end{table}

In Table \ref{t:spm}, the base station antenna array height for the
cellular system is denoted by $h_{\sst{BS}}$. Taking into account that
it is more convenient and cost effective to place all the service
antennas in the same tower, we assume a higher antenna tower for
cellular case. The valid range for $d$ is between 10 m and 5000 m. In
our scenarios, we always have $d>10$ because all the service antennas
are placed more than 10 meters higher than the user antenna.

\subsubsection{Coverage Region}

\begin{figure}
\centering
\includegraphics[width=3.1in]{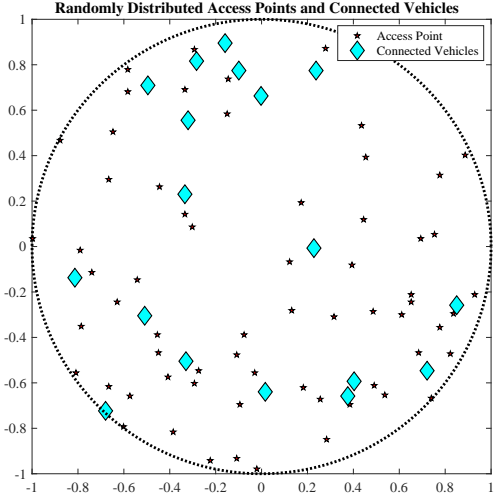}
\caption{An example of cell-free Massive MIMO: $M=64$ access points
  and $K=18$ connected vehicles. A corresponding cellular Massive MIMO
  serves the same circular area with a base station with an $M$
  antenna array at the center.}
\label{f:cf_cir}
\end{figure}

The service region is a circular disk with radius $R$. For the
cellular system, a Massive MIMO base station with $M$ service antennas
is located at the center of the circle, and $K$ autonomous vehicles
are randomly distributed inside the circle. For the cell-free system, $M$
single-antenna access points and $K$ autonomous vehicles are randomly
distributed inside a circle.  Fig. \ref{f:cf_cir} depicts an example
of the cell-free system. In both systems, vehicles near the center of
the service area are statistically different from those near the edge.

\subsection{Single Cell with MR Decoding}

The uplink effective SINR for single-cell Massive MIMO with MR decoding is given by \cite{mlyn_2016}
\begin{equation}\label{eq:clmr}
\mbox{SINR}_k^{\sst{cl,MR}}={M\rhou\gamma_k\eta_k\over 1+\rhou\sum_{k'=1}^K\beta_{k'}\eta_{k'}}, \quad k=1,\cdots, K
\end{equation}
where 
\begin{equation}\label{eq:gamma}
\gamma_k={\rhou\tau\beta_k^2\over 1+\rhou\tau\beta_k}
\end{equation}
is the mean-square of the channel estimate, and $\boldsymbol\eta$ is the uplink power control with individual constraint expressed in  (\ref{eq:eta}).

The max-min power control that maximize the minimum uplink SINR can be easily obtained \cite{mlyn_2016}. The resulting maximal common SINR that is achieved by all users can be written explicitly as
$$
\mbox{SINR}^{\sst{cl,MR}}_{\sst{mm}}={M\rhou\over {1\over \min_{k'}\{\gamma_{k'}\}} + \rhou\sum_{k'=1}^K{\beta_{k'}\over \gamma_{k'}}}
$$
The max-min power control is given by 
\begin{equation}\label{eq:clmm}
\eta_k={\min_{k'}\{\gamma_{k'}\}\over \gamma_k}, \quad k = 1, \cdots, K.
\end{equation}

The top table in Table \ref{t:cl} tabulates the SE (spectral efficiency) performance
of single cell Massive MIMO with MR decoding. It is known \cite{ym_2014} that due to the severe near-far
problem in cellular networks, full power strategy (i.e.,
each user terminal transmits full power) with MR decoding does not work. However, max-min
power control can be used to achieve the 6 bps/Hz target with high probability. Note that
suburban and rural scenarios require very large numbers of service
antennas due to the large path losses. Furthermore, the suburban and
rural SE numbers (in \textcolor{red}{red} with an asterisk)
are 95\% likely instead of 99\% likely as in the urban case.

\begin{remark}
To achieve 6 bps/Hz spectral efficiency, we need an SINR of $2^6 -1 = 63
\approx 18$ dB. It has been shown \cite{mlyn_2016} that for any power
control, the arithmetic mean of the effective SINR given in
(\ref{eq:clmr}) over $K$ users is upper bounded by $M/K$. Thus for
single cell MR decoding we must have $M\ge 63K=63\times
18=1134$. The top table in Table \ref{t:cl} shows that we need many more than 1134
antennas to support suburban and rural scenarios.\qed
\end{remark}

\subsection{Single Cell with ZF Decoding}

The uplink effective SINR for single-cell Massive MIMO with ZF
decoding is given by \cite{mlyn_2016}
$$
\mbox{SINR}_k^{\sst{cl,ZF}}={(M-K)\rhou\gamma_k\eta_k\over 1+\rhou\sum_{k'=1}^K(\beta_{k'}-\gamma_{k'})\eta_{k'}}, \quad k=1,\cdots, K
$$
where $\gamma_k$ is given by (\ref{eq:gamma}). 

The maximal common SINR that can be achieved by all users is
$$
\mbox{SINR}^{\sst{cl,ZF}}_{\sst{mm}}={(M-K)\rhou\over {1\over \min_{k'}\{\gamma_{k'}\}} + \rhou\sum_{k'=1}^K{\beta_{k'} - \gamma_{k'}\over \gamma_{k'}}}.
$$ The corresponding max-min power control is given by (\ref{eq:clmm})
also.

The bottom table in Table \ref{t:cl} tabulates the spectral efficiency performance of
single cell Massive MIMO with ZF decoding. The severe near-far
problems in cellular network are effectively mitigated by ZF decoding.
We see that full power strategy actually outperforms max-min power
control in terms of 99\% likely throughput (for urban) and 95\% likely
throughput (for suburban and rural, indicated by \textcolor{red}{red}
numbers with asterisk). Remark \ref{r:rate} below explains how this is statistically possible. Note that suburban and rural scenarios still
require large numbers of service antennas due to the large path losses.

\begin{table}
\caption{Cellular, 99\% Likely SE (bps/Hz), $K=18$.}\label{t:cl}
\centering
\begin{tabular}{|c|c|c|c|c|} \hline
& & Full Power & Max-Min & $M$ \\ \hline
\multirow{3}{*}{\rotatebox[origin=c]{20}{Maximum Ratio}} & Urban & 0.05  & 6.1 &  1700 \\ \cline{2-5} 
& Suburban & \textcolor{red}{$0.3^*$} & \textcolor{red}{$6^*$} &  16000 \\ \cline{2-5}
& Rural & \textcolor{red}{$1.5^*$}  & \textcolor{red}{$6^*$} &  378000 \\ \hline\hline
\multirow{3}{*}{\rotatebox[origin=c]{20}{Zero-Forcing}} & Urban & 6.1  & 4.8 &  200 \\ \cline{2-5}
& Suburban & \textcolor{red}{$6^*$}  & \textcolor{red}{$2.2^*$} & 1000  \\ \cline{2-5}
& Rural & \textcolor{red}{$6.1^*$}  & \textcolor{red}{$1.6^*$} & 11000  \\ \hline
\end{tabular}
\end{table}

\subsection{Cell-Free with MR Decoding}

The uplink effective SINR for cell-free Massive MIMO with MR decoding is given by \cite{ngo_2017}
\begin{eqnarray}
\mbox{SINR}_k^{\sst{cf,MR}}={\rhou\displaystyle\left(\sum_{m=1}^M\gamma_{m,k}\right)^2\eta_k\over \displaystyle\sum_{m=1}^M\gamma_{m,k}+\rhou\sum_{k'=1}^K\eta_{k'}\sum_{m=1}^M\gamma_{m,k}\beta_{m,k'}}, \label{eq:sinrcfmr}\\
\hspace{2in} k=1,\cdots, K \nonumber
\end{eqnarray}
where 
\begin{equation}\label{eq:gmk}
\gamma_{m,k}={\rhou\tau\beta_{m,k}^2\over 1+\rhou\tau\beta_{m,k}}
\end{equation}
is the mean-square of the channel estimate. 

\begin{remark}
It is known \cite{caire_2018,cb_2018} that the SINR given by
(\ref{eq:sinrcfmr}) is in general too conservative, due to slower
channel hardening in cell-free Massive MIMO. Yet, the 99\% likely data
rates given by (\ref{eq:sinrcfmr}) are reasonably accurate, as shown
in \cite{cb_2018}. \qed
\end{remark}

Let ${\mathbb R}^M_{+0}$ denote the set of $M$ dimensional real vectors with non-negative entries, and
\begin{eqnarray*}
\boldsymbol\gamma_k&&\hspace{-.25in}=\left[\gamma_{1,k}, \cdots, \gamma_{M,k}\right]^{\sst T}\in \mathbb R^M_{+0} \\
\boldsymbol\beta_k&&\hspace{-.25in}=\left[\beta_{1,k}, \cdots, \beta_{M,k}\right]^{\sst T}\in \mathbb R^M_{+0} \\
{\bf 1}_M&&\hspace{-.25in}=\left[ 1, \cdots, 1\right]^{\sst T}\in \mathbb R^M_{+0}
\end{eqnarray*}

The max-min power control and the achieved common maximum SINR can be
obtained by using bisection to search for the largest $\zeta$ that
satisfies linear equation (\ref{eq:leq}), with solution
$\boldsymbol\eta$ satisfying (\ref{eq:eta}).
\begin{eqnarray}
\lefteqn{\left(
\left[
\begin{array}{ccc}
<\boldsymbol\gamma_1,{\bf 1}_M>^2 & \cdots & 0\\
\vdots & \ddots & \vdots \\
0 & \cdots & <\boldsymbol\gamma_K,{\bf 1}_M>^2 
\end{array}
\right]- \right.} \nonumber\\
&& \left. \zeta 
\left[
\begin{array}{ccc}
<\boldsymbol\gamma_1,\boldsymbol\beta_1> & \cdots & <\boldsymbol\gamma_1,\boldsymbol\beta_K>\\
\vdots & \ddots & \vdots \\
<\boldsymbol\gamma_K,\boldsymbol\beta_1> & \cdots & <\boldsymbol\gamma_K,\boldsymbol\beta_K> 
\end{array}
\right]\right)\boldsymbol\eta  \nonumber \\
&& \hspace{1in}=\left({\zeta\over \rhou}\right)\left[
\begin{array}{c}
<\boldsymbol\gamma_1,{\bf 1}_M> \\
\vdots \\
<\boldsymbol\gamma_K,{\bf 1}_M>
\end{array}
\right]  \label{eq:leq}
\end{eqnarray}

From (\ref{eq:sinrcfmr}), regardless what power control strategy is
used, we can upper bound the effective SINR for the $k$th terminal as
follows:
\begin{eqnarray}
\mbox{SINR}_k^{\sst{cf,MR}}&&\hspace{-.25in}<{\displaystyle\left(\sum_{m=1}^M\gamma_{m,k}\right)^2\eta_k\over \displaystyle\sum_{k'=1}^K\eta_{k'}\sum_{m=1}^M\gamma_{m,k}\beta_{m,k'}} \nonumber \\
&&\hspace{-.25in}\le {\displaystyle\left(\sum_{m=1}^M\gamma_{m,k}\right)^2\over \displaystyle\sum_{m=1}^M\gamma_{m,k}\beta_{m,k}}\nonumber \\
&&\hspace{-.25in}<{\displaystyle\left(\sum_{m=1}^M\gamma_{m,k}\right)^2\over \displaystyle\sum_{m=1}^M\gamma_{m,k}^2}  =\left< {\boldsymbol\gamma_k\over \|\boldsymbol\gamma_k\|_2}, {\bf 1}_M\right>^2  \label{eq:ub}
\end{eqnarray}

In particular, for max-min power control, the achieved common SINR for all $K$ user terminals is upper bounded by
\begin{equation}\label{eq:ubmm}
\min_{k=1,\cdots, K} \left< {\boldsymbol\gamma_k\over \|\boldsymbol\gamma_k\|_2}, {\bf 1}_M\right>^2
\end{equation}

\begin{lemma}\label{l:1}
Let ${\bf x}=\left[ x_1, \cdots, x_M\right]^{\sst T}\in {\mathbb R}^M_{+0}$. Then
\begin{enumerate}
\item
$$
\displaystyle\max_{{\bf x}\in {\mathbb R}^M_{+0}, \|{\bf x}\|_2 =1}\left<{\bf x}, {\bf 1}_M\right>=\sqrt{M}
$$
and 
$
\left<{\bf x}, {\bf 1}_M\right>=\sqrt{M} 
$
if and only if $x_1=\cdots = x_M$.
\item
$$
\displaystyle\min_{{\bf x}\in {\mathbb R}^M_{+0}, \|{\bf x}\|_2 =1}\left<{\bf x}, {\bf 1}_M\right>=1
$$
and
$\left<{\bf x}, {\bf 1}_M\right>=1$  if and only if there exists an entry $x_i$ of ${\bf x}$ such that $x_i=1$ and $x_j=0\quad \forall j\ne i$.
\end{enumerate}
\end{lemma}

{\it Proof:}
1) is obtained by the Cauchy-Schwarz inequality. 

To prove 2), noting that $\sum_{m=1}^M x_m^2=\|{\bf x}\|_2^2=1$, we have
$x_m\le 1 \quad \forall m$. Thus $\left<{\bf x}, {\bf 1}_M\right>=\sum_{m=1}^M x_m \ge \sum_{m=1}^M x_m^2=1$. 

If two or more entries of ${\bf x}$ are greater than zero, then
$\left( \sum_{m=1}^M x_m\right)^2 > \sum_{m=1}^M x_m^2=1$. Thus only one entry of ${\bf x}$ can be nonzero if $\left<{\bf x}, {\bf 1}_M\right>=1$. The nonzero entry must be 1 since $\|{\bf x}\|_2=1$.
\qed

By Lemma \ref{l:1}, the upper bound (\ref{eq:ub}) is maximized ($=M$) if and only if
$$
\gamma_{1,k}=\cdots = \gamma_{M,k},
$$
which is precisely the single cell case where $M$ access point
antennas are co-located. In the cell-free case, this bound can often be
very close to 1, which is the minimum for this upper bound by Lemma
\ref{l:1}, in the case when there is a dominant access point with
large $\beta$, and the rest of the $\beta$'s are much smaller
relatively.  This means that the uplink SINR for the cell-free Massive
MIMO with MR decoding is not scalable: the upper bound for the SINR
can be very close to 1, regardless how large $M$ is.

\begin{figure}
\centering
\includegraphics[width=3.5in]{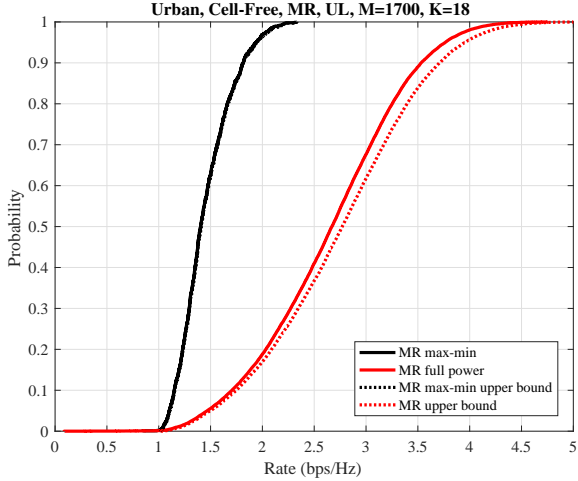}
\caption{Per vehicle uplink data rate. Cell-free with MR decoding in urban scenario: 1700 antennas serve 18 vehicles inside the circle with 0.5 km radius.}
\label{f:urbancfmr}
\end{figure}

Figure \ref{f:urbancfmr} shows the CDF (cumulative distribution
function) of the per-vehicle uplink rate in the urban scenario. Even
though 1700 AP's are employed, there is a non-negligible probability
that the SINR is close to 1, which corresponds to a spectral
efficiency of $\log_2(1+1)=1$ bps/Hz. It is impossible to achieve a
99\%-likely per user rate of 6 bps/Hz even with a huge $M$. A similar
behavior is also observed in suburban and rural scenarios. It is
interesting to note that the upper bounds (\ref{eq:ub}) and
(\ref{eq:ubmm}) are remarkably tight in Figure \ref{f:urbancfmr}, and
as $M$ increases, these upper bounds virtually overlap the
corresponding CDF curves in suburban and rural scenarios. Note that no
power control strategy can do better than the ``MR upper bound'', but
the full-power strategy comes close.

The top table in Table \ref{t:cf} tabulates the spectral efficiency performance of
cell-free Massive MIMO with MR decoding. Same as in Table
\ref{t:cl}, the \textcolor{red}{red} numbers with
asterisk are the 95\% likely rates. Since the effective SINR's are
upper bounded by (\ref{eq:ub}) and (\ref{eq:ubmm}), supporting uplink
intensive applications like connected vehicles with MR decoding in
cell-free configuration is, unfortunately, not possible.

\begin{remark}\label{r:rate}
Note that Figure \ref{f:urbancfmr} and Table \ref{t:cf} show that full power strategy gives better 99\%-likely per vehicle rate than max-min power control. This can be explained as follows.   
We use 1000 independent realizations of large-scale fading profiles to produce Figure \ref{f:urbancfmr}. Thus each CDF curve is generated with $1000\times 18 = 18000$ data points. Within each realization of large-scale fading, at least one of the vehicles with full power strategy will have a lower rate than all the vehicles with max-min power control, which imposes the same rate to all vehicles. But often the lowest rate for full power strategy in one large-scale fading realization is higher than the max-min rates for many other large-scale fading realizations. It is theoretically possible that among 18000 rates from full power control strategy, only one is lower than all 1000 max-min rates.  \qed
\end{remark}

\begin{remark}
We note that downlink performance for cell-free Massive MIMO with MR
precoding \cite{ngo_2017,elina_2017} is not confined by upper
bounds similar to (\ref{eq:ub}) and (\ref{eq:ubmm}) because each
access point can allocate different powers to different users, and
thereby better control the interference. Note that the downlink power
control has $MK$ tunable coefficients while uplink power control has
only $K$. \qed
\end{remark}

\subsection{Cell-Free with ZF Decoding}

Saving the best for last, we next consider cell-free Massive MIMO with
ZF decoding, which has not been studied in the existing literature.

We shall first derive an expression for the uplink SINR. 

\subsubsection{Uplink Effective SINR} 

The ZF decoding matrix is
$$
A_{\sst{ZF}}=(\hat G^*\hat G)^{-1}\hat G^*,
$$
where $\hat G$ is given by (\ref{eq:ghat}), and the superscript $\phantom{a}^*$ denotes conjugate transpose.
Applying ZF decoding we have
\begin{eqnarray*}
A_{\sst{ZF}}{\bf y}\hspace{-.25in}&&=\sqrt{\rhou}A_{\sst{ZF}}G{\bf x} + A_{\sst{ZF}}{\bf w} \\
\hspace{-.25in}&&=\sqrt{\rhou}A_{\sst{ZF}}(\hat G + \tilde G){\bf x} + A_{\sst{ZF}}{\bf w} \\
\hspace{-.25in}&&=\sqrt{\rhou}{\bf x} +\sqrt{\rhou}A_{\sst{ZF}}\tilde G{\bf x} + A_{\sst{ZF}}{\bf w},
\end{eqnarray*}
where $\tilde G$ is given by (\ref{eq:gtilde}).

The signal power for the $k$th terminal is given by
$$
\rhou{\text E}x_k^*x_k = \rhou\eta_k.
$$
The effective interference power of the system is contributed by the term
$$
\sqrt{\rhou}A_{\sst{ZF}}\tilde G{\bf x} + A_{\sst{ZF}}{\bf w}
$$ which consists of interference due to channel estimation error and
noise. To compute the effective interference power, we first compute
the covariance
\begin{eqnarray*}
\lefteqn{{\text{Cov}}(\sqrt{\rhou}A_{\sst{ZF}}\tilde G{\bf x} + A_{\sst{ZF}}{\bf w}) =} \\
&& = {\text E}(\sqrt{\rhou}A_{\sst{ZF}}\tilde G{\bf x} + A_{\sst{ZF}}{\bf w}) (\sqrt{\rhou}A_{\sst{ZF}}\tilde G{\bf x} + A_{\sst{ZF}}{\bf w})^* \\
&& = \rhou{\text E}(A_{\sst{ZF}}\tilde G{\bf x}{\bf x}^*\tilde G^*A_{\sst{ZF}}^*)+{\text E}(A_{\sst{ZF}}{\bf w}{\bf w}^*A_{\sst{ZF}}^* )\\
&& = \rhou{\text E}\{A_{\sst{ZF}}\tilde G[{\text{diag}}(\eta_1, \cdots, \eta_K)]\tilde G^*A_{\sst{ZF}}^*\}+{\text E}(A_{\sst{ZF}}A_{\sst{ZF}}^*)\\
&& = \rhou{\text E}\left\{A_{\sst{ZF}}\left[\sum_{k'=1}^K (\tilde{\bf g}_{k'}\tilde {\bf g}_{k'}^*)\eta_{k'}\right]A_{\sst{ZF}}^*\right\}+{\text E}(\hat G^*\hat G)^{-1}
\end{eqnarray*}
where $\tilde {\bf g}_{k'}$ is the $k'$th column of the estimation error $\tilde G$.

The effective interference power for the $k$th terminal is the $k$th diagonal element of the covariance matrix
$$
\rhou\left[{\text E}\left\{A_{\sst{ZF}}\left[\sum_{k'=1}^K (\tilde{\bf g}_{k'}\tilde {\bf g}_{k'}^*)\eta_{k'}\right]A_{\sst{ZF}}^*\right\}\right]_{k,k}+\left[{\text E}(\hat G^*\hat G)^{-1}\right]_{k,k}.
$$
Here $[A]_{k,k}$ is the $k$th diagonal element of the matrix $A$.

The effective uplink SINR for the $k$th user is then given by
\begin{eqnarray*}
\lefteqn{{\mbox{SINR}}_k^{\sst{cf,ZF}}=} \\
&&\hspace{-.25in}{\rhou\eta_k\over \displaystyle\rhou\left[{\text E}\left\{A_{\sst{ZF}}\left[\sum_{k'=1}^K (\tilde{\bf g}_{k'}\tilde {\bf g}_{k'}^*)\eta_{k'}\right]A_{\sst{ZF}}^*\right\}\right]_{k,k}+\left[{\text E}(\hat G^*\hat G)^{-1}\right]_{k,k}}
\end{eqnarray*}

With some algebra, we can rewrite the interference power due to channel estimation error as
$$
\rhou\sum_{k'=1}^K \eta_{k'}{\text E}(b_{k,k'}^*b_{k,k'})=\rhou\sum_{k'=1}^K \eta_{k'}{\text E}\left(|b_{k,k'}|^2\right),
$$
where 
$
[b_{i,j}]=B\triangleq A_{\sst{ZF}}\tilde G.
$

Then the effective uplink SINR for the $k$th user is given by
\begin{equation}\label{eq:sinr}
{\mbox{SINR}}_k^{\sst{cf,ZF}}=
{\rhou\eta_k\over \displaystyle\rhou\sum_{k'=1}^K \eta_{k'}{\text E}\left(|b_{k,k'}|^2\right)+\left[{\text E}(\hat G^*\hat G)^{-1}\right]_{k,k}}  
\end{equation}

We next provide formulas for calculating the max-min power control and the corresponding SINR.

\subsubsection{Max-Min Power Control}\label{s:cfzfmm}

Let $\zeta$ be the common SINR achieved by the $K$ terminals. From (\ref{eq:sinr}) we have 
\begin{equation}\label{eq:sinr1}
{\rhou\eta_k\over \displaystyle\rhou\sum_{k'=1}^K \eta_{k'}{\text E}\left(|b_{k,k'}|^2\right)+\left[{\text E}(\hat G^*\hat G)^{-1}\right]_{k,k}}\equiv \zeta.
\end{equation}
We can rewrite (\ref{eq:sinr1}) in matrix form as
\begin{equation}\label{eq:sinr2}
(I_K-\zeta {\text E}(B^*\circ B))\boldsymbol\eta = {\zeta\over \rhou}{\text{diag}}[{\text E}(\hat G^*\hat G)^{-1})],
\end{equation}
where $I_K$ is the $K$-dimensional identity matrix, and $B^*\circ B$
is the element-wise multiplication (Hadamard product),
${\text{diag}}[{\text E}(\hat G^*\hat G)^{-1})]$ is the vector formed
by the diagonal elements of ${\text E}(\hat G^*\hat G)^{-1}$.

Bisection search is used to obtain maximum $\zeta$ that satisfies (\ref{eq:sinr2}) and the corresponding power control $\boldsymbol\eta$ that satisfies (\ref{eq:eta}). 

\begin{figure}
\centering
\includegraphics[width=3.5in]{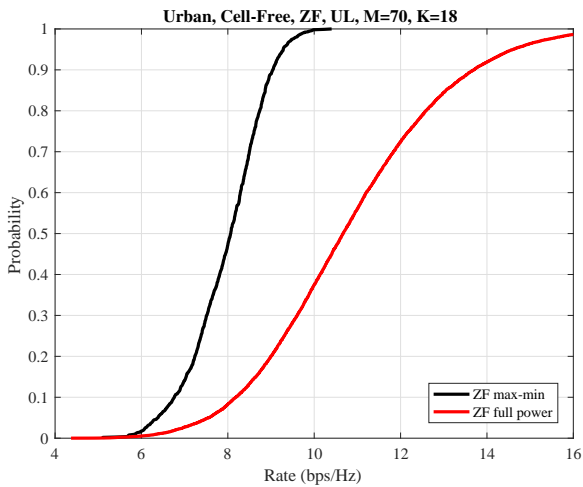}
\caption{Per vehicle uplink data rate. Cell-free with ZF decoding in urban scenario: 70 antennas serve 18 vehicles inside the circle with 0.5 km radius.}
\label{f:urbancfzf}
\end{figure}

Fig. \ref{f:urbancfzf} shows the CDF of the
per vehicle uplink rate in the urban scenario. It clearly shows that full power strategy outperforms max-min power control. Similar CDF's are
observed for the suburban and rural scenarios. 

The bottom table in Table \ref{t:cf} summarizes the performance of cell-free Massive
MIMO with ZF decoding. The full power strategy outperforms max-min
power control in terms of 99\% likely throughput (see Remark \ref{r:rate} for how this is statistically possible). For urban
propagation within a 0.5 km radius, 70 randomly placed access points
are enough to support 18 autonomous vehicles; for suburban propagation
within 1 km radius, and for rural 450 MHz band propagation within 4 km radius, 100
access points are enough. 

\begin{table}
\centering
\caption{Cell-Free, 99\% Likely SE (bps/Hz), $K=18$.}\label{t:cf}
\begin{tabular}{|c|c|c|c|c|} \hline
& &  Full Power &  Max-Min &  $M$  \\ \hline
\multirow{3}{*}{\rotatebox[origin=c]{20}{Maximum Ratio}} & Urban & 1.2  & 1 & 1700  \\ \cline{2-5} 
& Suburban & 1.2 (\textcolor{red}{$1.6^*$})  & 1 (\textcolor{red}{$1.1^*$})  & 16000  \\ \cline{2-5}
& Rural & 1.7 (\textcolor{red}{$2.5^*$})  & 1.2 (\textcolor{red}{$1.4^*$}) & 378000  \\ \hline \hline
\multirow{3}{*}{\rotatebox[origin=c]{20}{Zero-Forcing}} & Urban & 6.4  & 5.9  & 70  \\ \cline{2-5} 
& Suburban & 6.4 & 5.5  & 100  \\ \cline{2-5}
& Rural & 6.5  & 5.8  & 100   \\ \hline
\end{tabular}
\end{table}

\section{Conclusions}

New mobile applications such as connected vehicles (cars, drones)
demand high uplink throughput, low latency, and high reliability. Due
to the limited availability of bandwidth in the frequency band that is
suitable for mobile applications, the required SINR is high. Meeting
such demands may appear to be extremely challenging for a wireless
network.

A pleasant surprise is that cell-free Massive MIMO with ZF decoding
can easily support such applications in urban, suburban and rural
morphologies. An important added advantage is that no power control is
needed: every vehicle just transmits with full power. Employing
cellular Massive MIMO with ZF decoding to support such applications is
also possible if the cell size is small, such as in an urban
morphology. Because high SINR is required, using Massive MIMO with MR
decoding is in general not desirable. Indeed, our simulations show
that MR decoding is not suitable for supporting such applications,
either cellular or cell-free. A rather unpleasant surprise is that in
terms of 99\% likely user uplink throughput, cell-free Massive MIMO
with MR decoding severely underperforms cellular due to strong
self-interference.

By considering only the single cell case, we overestimate the cellular
performance. Yet results still indicate that a cellular network is not
suitable for supporting uplink-intensive applications. We only
consider single-antenna terminals. Multiple antenna terminals can be
deployed to further increase throughput.

Further investigations should include the performance of MMSE (minimum mean square error) type receivers and maximum-ratio combination with weights \cite{elina_2016}, and correlated and Ricean channels \cite{ngo_2018}.

\section*{Acknowledgment}
The authors would like to thank Prof. Thomas L. Marzetta for his helpful comments.

\end{document}